\documentclass{article}

\usepackage[english]{babel}

\usepackage[a4paper,top=2cm,bottom=2cm,left=3cm,right=3cm,marginparwidth=1.75cm]{geometry}

\usepackage{tabularx}
\usepackage{amssymb}
\usepackage{siunitx}
\PassOptionsToPackage{hyphens}{url}\usepackage{hyperref}
\usepackage{cleveref}
\usepackage[utf8]{inputenc}
\usepackage[right]{lineno}
\usepackage{csquotes}
\usepackage{booktabs}
\usepackage{longtable}
\usepackage{adjustbox}
\usepackage{array}
\usepackage{url}
\usepackage{titlesec}
\usepackage{authblk}
\usepackage{float}
\usepackage{wrapfig}
\usepackage{listings}
\usepackage{xcolor}
\usepackage{fvextra}
\usepackage{tcolorbox}
\usepackage{xcolor}

\newcommand{\rev}[1]{{#1}} 

\titleformat{\subsection}
  {\mdseries\itshape\large} 
  {\thesubsection}{1em}{} 

\usepackage[numbers]{natbib}
\crefformat{figure}{#2Figure~#1#3}
\Crefformat{figure}{#2Figure~#1#3}
\crefformat{table}{#2Table~#1#3}
\Crefformat{table}{#2Table~#1#3}
\crefformat{section}{#2Section~#1#3}
\Crefformat{section}{#2Section~#1#3}

\author[1]{John Wu}
\author[1]{Zhenbang Wu}
\author[1,*]{Jimeng Sun}
\affil[1]{University of Illinois, Urbana-Champaign, USA}
\affil[*]{\rev{Corresponding author: \texttt{jimeng@illinois.edu}}}

\title{Bridging the Reproducibility Divide: Open Source Software's Role in Standardizing Healthcare AI}

\begin{document}
\date{}
\maketitle

\begin{abstract}
Artificial intelligence for healthcare (AI4H) has significant health, ethical, and legal implications due to its direct impact on human lives. However, the field is facing a reproducibility crisis that raises concerns about the deployment of AI4H systems.
Our analysis of recent AI4H publications reveals that, despite a trend toward utilizing open datasets and sharing modeling code, 74\% of AI4H papers still rely on private datasets or do not share their code. This is especially concerning in healthcare applications, where trust is essential. Furthermore, inconsistent and poorly documented data preprocessing pipelines result in variable model performance reports, even for identical tasks and datasets, making it challenging to evaluate the true effectiveness of AI models.

Despite the challenges posed by the reproducibility crisis, addressing these issues through open practices offers substantial benefits. For instance, while the reproducibility mandate adds extra effort to research and publication, it significantly enhances the impact of the work. Our analysis shows that papers that used both public datasets and shared code received, on average, 110\% more citations than those that do neither—more than doubling the citation count.

Given the clear benefits of enhancing reproducibility, it is imperative for the AI4H community to take concrete steps to overcome existing barriers. The community should promote open science practices, establish standardized guidelines for data preprocessing, and develop robust benchmarks. Tackling these challenges through open-source development can improve reproducibility, which is essential for ensuring that AI models are safe, effective, and beneficial for patient care. This approach will help build more trustworthy AI systems that can be integrated into healthcare settings, ultimately contributing to better patient outcomes and advancing the field of medicine.
\end{abstract}

\section{Introduction} 
AI models are increasingly used in healthcare for tasks such as disease diagnosis, drug recommendation, medical coding, report generation, and mortality prediction \cite{Habehh2021-tn_MLHC_survey}. However, improper validation and deployment of these models can lead to severe consequences, ranging from inadequate coverage of patients \cite{mittermaier2023bias_in_medical_AI} to legal challenges \cite{napolitano_2023_health_insur_lawsuit, vcartolovni2022ethical_legal_considerations_review}. 
The FDA's regulatory framework for AI/ML-based Software as a Medical Device (SaMD) has approved over 500 AI-enabled medical devices as of 2024, with the approval process heavily emphasizing model validation and reproducibility \cite{Center-for-Devices2024-zp}. The FDA's {\it predetermined change control plan} specifically requires manufacturers to demonstrate evidence-based and transparent approaches with reproducible model performance~\cite{Center-for-Devices2023-ee}.
Given the high-stakes nature of healthcare \cite{10.1371/journal.pmed.0020124_whymostresearchfalse}, validating and reproducing research findings is crucial not only for advancing state-of-the-art research in AI4H but also for ensuring the safe deployment of these models in clinical practice.

This paper aims to define reproducibility in the context of AI4H in 2024, describe its research landscape, and propose practical steps to address the reproducibility challenges and improve transparency. To the best of our knowledge, our study presents the first large-scale analysis of reproducibility in AI4H as of 2024, examining thousands of papers. While \cite{mcdermott2019reproducibilitymachinelearninghealthcrisis} provides a highly detailed analysis of AI4H in 2018, our work expands the scope by an order of magnitude and introduces an automated approach to assess reproducibility, which is particularly valuable given the significant increase in the number of AI4H publications. Unlike general AI domains studied in prior reproducibility work \cite{gundersen2018reproducible_call, hutson2018artificial_crisis_review_call}, healthcare presents distinct challenges with its high-stakes outcomes, privacy constraints, and need for domain expertise. Furthermore, our analysis incorporates additional metrics such as citation patterns and research topics to demonstrate the tangible benefits of enhancing reproducibility. As the AI4H field has expanded tremendously in the past five years \cite{senthil2024bibliometric_trends_AI4H}, we believe it is critical for another review of the state of reproducibility in AI4H.

\begin{figure}
    \centering
    \includegraphics[width=1.0\linewidth]{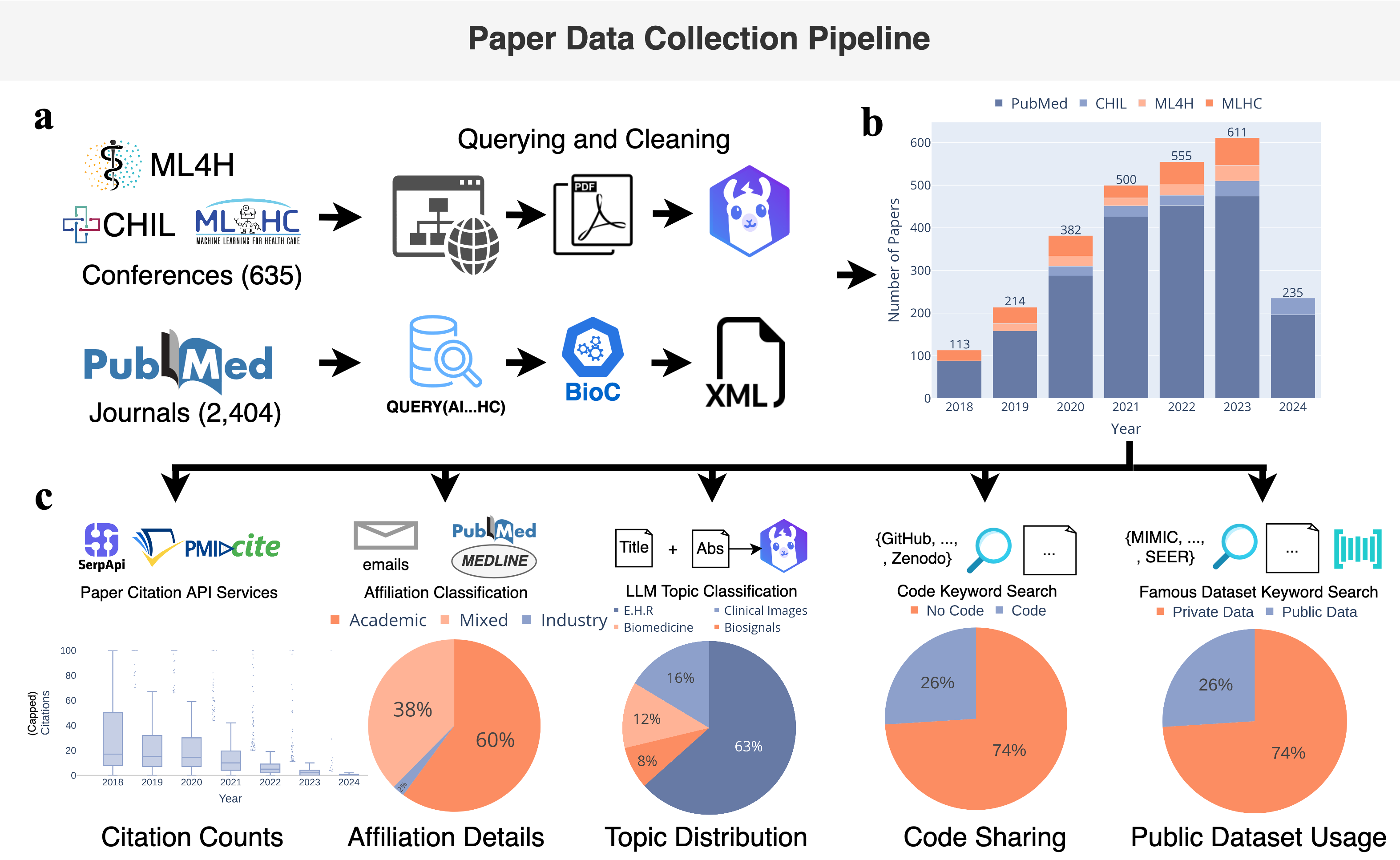}
    \caption{Paper Data Collection and Analysis: (a) Paper scraping details: We scrape PDF files directly from conference webpages, extracting and cleaning important paper details like title, authors, abstracts, and emails. We query BioC to scrape paper information from PubMed's Open Access Database, related to research papers that contain the terms AI, deep learning, machine learning, healthcare, electronic health records, and electronic medical records into XML formats. (b) Stacked bar chart of total number of papers scraped each year, showing a steady increase in paper counts over time. (c) Key analyses performed: Mapping citation counts to each paper using scholarly services like SerpAPI \cite{serpapi2024}, Semantic Scholar \cite{Kinney2023TheSS_Semantic_scholar}, and PMIDcite \cite{gusenbauer2020academic_pmid_cite}; scraping affiliation details by classifying emails or affiliations sourced from PubMed's MEDLINE API \cite{gusenbauer2020academic_pmid_cite}; classifying each paper's topic by its abstract and title with a medically fine-tuned large language model (OpenBioLLM-70b); checking for code sharing by identifying specific code keywords (e.g., GitHub, Zenodo, Colab, GitLab) in each paper's main text (excluding the references); assessing public dataset usage by checking for famous dataset keywords in the main text and cross-referencing each paper with the PapersWithCode API. We go into further detail in Appendix \ref{apd: Methodology Details}.}
    \label{fig:scraping}
\end{figure}

\subsection{Technical Reproducibility Is a Cornerstone in AI4H}
Reproducibility, due to its multi-faceted nature, encompasses a variety of definitions across different fields \cite{desai2024reproducibilityartificialintelligencemachine_definition, NAP25303_Broad_definitions_reproducibility}. In computer science, the Association for Computing Machinery (ACM) defines reproducibility as the ability of a different team to obtain the same results using the same experimental setup (i.e., code, data, and methodological specifics) \cite{semmelrock2023reproducibilitymachinelearningdrivenresearch}.  In the AI4H domain, McDermott et al. \cite{mcdermott2019reproducibilitymachinelearninghealthcrisis} further delineates reproducibility into three sub-categories:
\begin{enumerate}
    \item {\bf Technical reproducibility:} {\it Can results be reproduced under technically identical conditions?} 
    This involves reproducing the precise results reported in a study using the same code and dataset.
    \item {\bf Statistical reproducibility:} {\it Can results be reproduced under statistically identical conditions?} 
     This refers to reproducing statistically equivalent results when using resampled data or random seeds.
    \item {\bf Conceptual reproducibility:} {\it Can results be reproduced under conceptually identical conditions?} 
    This involves generalizing the study’s claims to new but conceptually identical settings (e.g., similar datasets, models, or tasks).
\end{enumerate}

The relationship between technical, statistical, and conceptual reproducibility in AI for healthcare is complex and multifaceted. Research has demonstrated that achieving technical reproducibility does not guarantee conceptual reproducibility, and furthermore, that conceptual reproducibility itself may be an inherently ambiguous and unnecessary objective. For instance, \cite{futoma2020myth_of_generalisability} presents a compelling argument that models can provide significant value even when their effectiveness is limited to specific patient populations. This perspective challenges the notion that broad generalizability is always necessary for clinical utility. In a related investigation, \cite{zech2018variable_generalisability} conducted an extensive study of deep learning models across diverse patient cohorts and healthcare systems. Their findings revealed that model performance is intrinsically linked to the characteristics of the training data distribution, and notably, that attempts to train models across multiple cohorts often result in the models inadvertently learning to exploit confounding variables - such as the originating hospital system - rather than clinically relevant features within medical imaging data. This phenomenon underscores the complexity of achieving true generalizability. Furthermore, in the domain of causal machine learning, performance heterogeneity across different populations is not viewed as a limitation but rather as a fundamental aspect of understanding treatment effects \cite{feuerriegel2024causal_ML}.

While these meta-analyses \cite{futoma2020myth_of_generalisability, zech2018variable_generalisability, feuerriegel2024causal_ML} present varying perspectives on conceptual reproducibility, they all implicitly depend on a foundation of technical reproducibility. The ability to replicate experiments and validate results across different patient populations serves as a crucial prerequisite for understanding how model performance varies across heterogeneous conditions. Without access to the underlying code and data, meaningful investigation of how new frontier AI models perform in diverse clinical contexts becomes virtually impossible. Moreover, the absence of technical reproducibility prevents direct verification of reported findings. This highlights a fundamental issue: while conceptual reproducibility or generalizability may neither be achievable nor necessary in all cases, technical reproducibility remains essential for rigorously examining and auditing claims about model performance and generalizability. Unfortunately, as illustrated in Figures \ref{fig:code_trends} and \ref{fig:data_trends}, achieving technical reproducibility continues to be a significant obstacle in the field of AI for healthcare.

Thus, this paper primarily focuses on the \textbf{technical reproducibility}, as it serves as a prerequisite to experimentally validating the statistical and conceptual reproducibility and is surprisingly a largely unresolved issue in the AI4H domain.

\section{The AI4H Reproducibility Crisis} \label{sec: Crisis}
To better understand the current reproducibility landscape, we scraped 528 papers from three major AI4H conferences, Conference on Health, Inference, and Learning (CHIL), Machine Learning for Health Symposium (ML4H), and Machine Learning for Healthcare (MLHC), as well as \rev{2,082} PubMed papers filtered by whether their manuscripts contained mentions of AI and healthcare from 2018 to 2024 (see Appendix for the detailed search query), as shown in Figure \ref{fig:scraping}. We investigated various aspects of technical reproducibility, such as public dataset usage and code sharing. 

We identify three key reproducibility challenges that uniquely persist in the AI4H domain:
\begin{enumerate}
\item \textbf{Private datasets:} Healthcare data's sensitive nature often necessitates the use of private datasets, preventing direct result replication by other researchers.
\item \textbf{Proprietary code:} Many AI4H models use proprietary algorithms, hindering transparency and reproducibility. 
\item \textbf{Lack of standardization in data processing:} Even when datasets and code are available, the absence of standardized preprocessing practices leads to significant variability in reported outcomes. 
\end{enumerate}

\subsection{Private Datasets}
\begin{figure}[t]
    \centering
\includegraphics[width=1.0\linewidth]{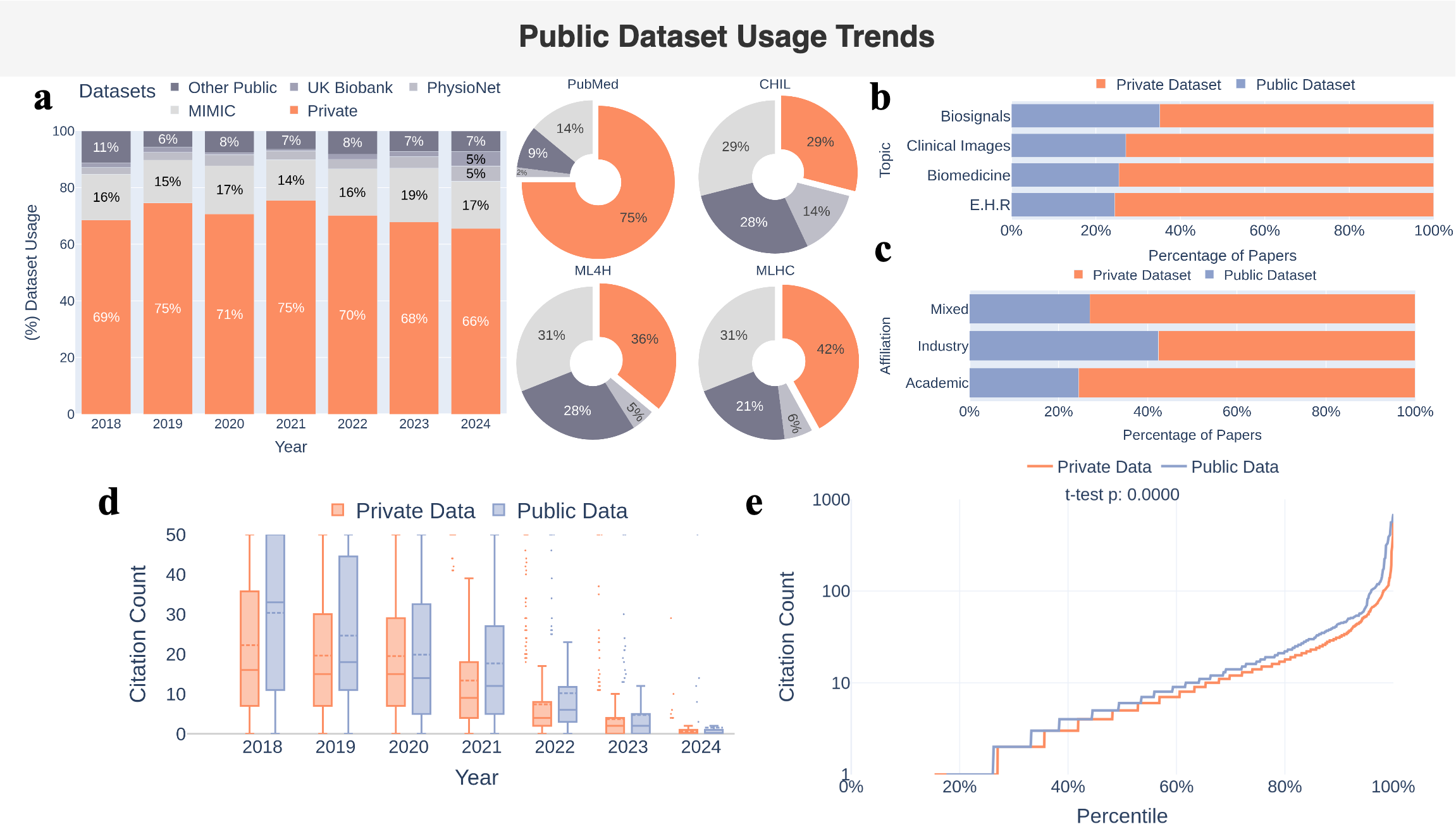}
    \caption{Trends in Public Dataset Usage: (a) Distribution of public dataset usage over time and across venues: MIMIC is the most commonly used or mentioned public dataset; private datasets dominate overall each year; conferences use more public datasets than journal papers. (b) Rate of public dataset usage across topics: Biosignal papers use the most public datasets. (c) Rate of public dataset usage across affiliations: Industry surprisingly uses the most public datasets. (d) Distribution of public dataset usage over time: Papers using public datasets have higher citation counts on average every year. The maximum number of citations on the y-axis is capped at 50 to focus on non-outlier behavior. (e) Cumulative distribution of using public datasets: Papers that use public datasets tend to have greater citation counts, regardless of outlier status.}
    \label{fig:data_trends}
\end{figure}

Private datasets pose a major obstacle to reproducibility in AI4H. Patient confidentiality and regulations like the Health Insurance Portability and Accountability Act (HIPAA) severely restrict access to raw patient data, preventing researchers from sharing datasets used in AI model development. These limitations make it nearly impossible to verify results from models trained on private data, such as Google's breast cancer detection model \cite{McKinney2020_google_breast_cancer_detection}. This is particularly concerning when models claim to outperform human experts. Furthermore, the variability in healthcare data across institutions and regions exacerbates the challenge of developing generalizable models.

Proposed alternatives have significant drawbacks. Synthetic datasets often lack the complexity of real patient data, omitting rare diseases, unique treatment combinations, and intricate demographic-health interactions. This can result in models that perform well in controlled settings but fail to generalize to real-world populations. De-identified datasets, while preserving clinical relevancy, risk re-identification through adversarial techniques \cite{Murdoch2021_privacy_concerns}. Additionally, over 89\% of patients oppose sharing their data for secondary use \cite{Murdoch2021_privacy_concerns}, further complicating dataset generation.


We analyzed mentions of well-known public dataset repositories in AI4H conferences and PubMed papers to estimate public dataset usage (Figure \ref{fig:data_trends}). It is important to note that this is a proxy measure and may not perfectly capture actual usage. Mentioning a public dataset name does not necessarily imply its use, and our approach might also overlook less well-known public datasets. The detailed validation is presented in Table \ref{tab:manual_validation}. 

Our analysis revealed several interesting trends across affiliations and topics. 
Figure~\ref{fig:data_trends}(a) confirmed that private datasets consistently dominate, accounting for roughly 65-75\% of all dataset usage from 2018 to 2024.
It also shows that 
those specialized AI healthcare conferences (CHIL, ML4H, and MLHC) tend to use more public datasets (about 60-70\%) compared to general medical literature in PubMed (only 25\%), suggesting these AI4H conferences might place greater emphasis on reproducibility and open science practices. 
This variance highlights the different data usage and sharing approaches within the AI4H community. Further details about the public datasets discussed in this analysis can be found in Appendix \ref{apd:public_dataset}.

As shown in Figure~\ref{fig:data_trends}(b-c), Biosignal papers contained the largest proportion of public dataset mentions, while papers authored solely by industry organizations mentioned public datasets more frequently than their academic or mixed counterparts. This surprising finding may be attributed to our classification of research hospitals as academic institutions, which are more likely to have access to private patient data than companies.

Another notable observation (Figure~\ref{fig:data_trends}(d-e)) was the correlation between public dataset mentions and higher citation counts, suggesting potential benefits for leveraging public datasets (t-test $p < 0.05$ in Figure \ref{fig:data_trends}e). 


\subsection{Proprietary Code}
\begin{figure}
    \centering
    \includegraphics[width=1.0\linewidth]{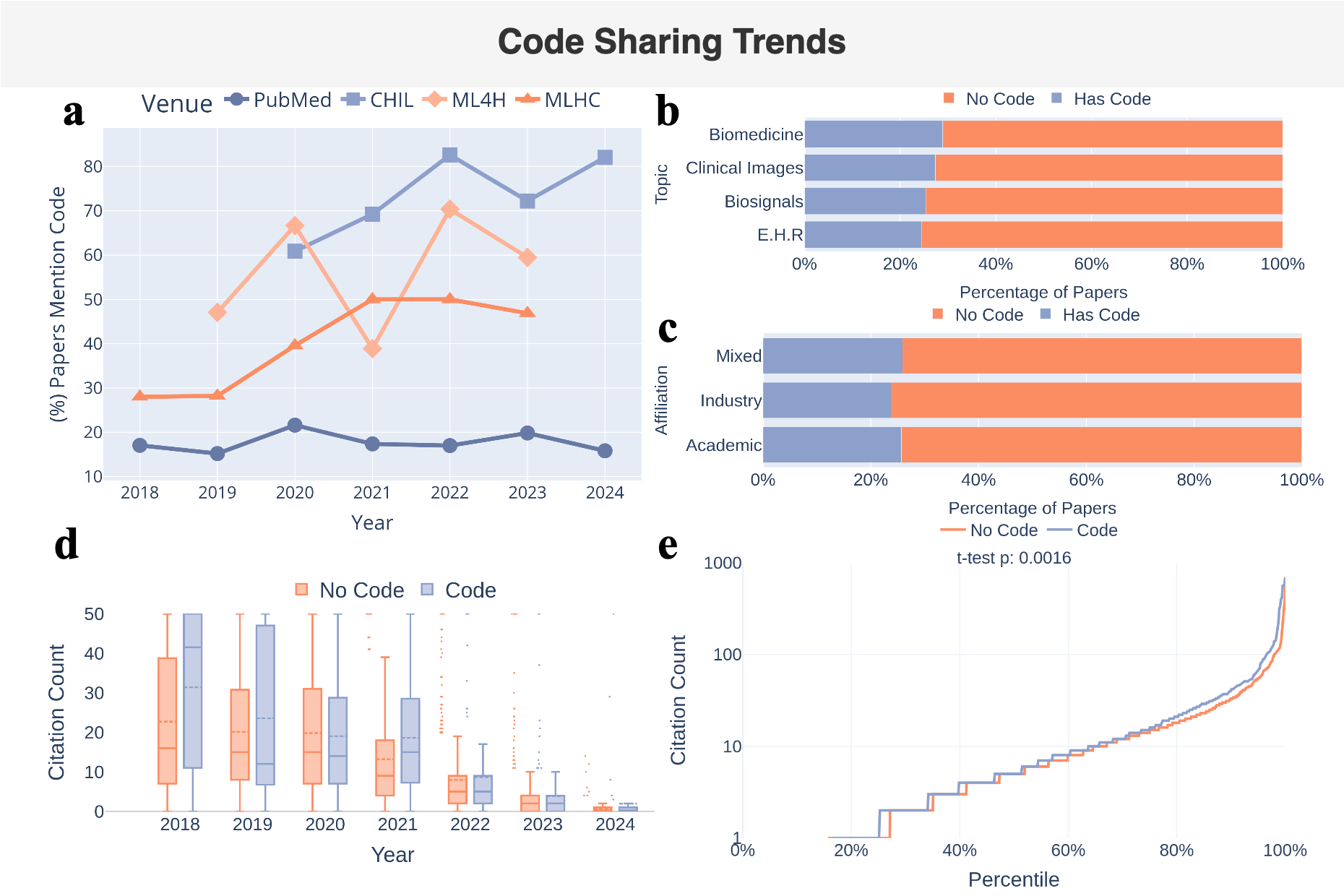}
    \caption{ Trends in Code Sharing: (a) Code-sharing percentage across different venues over time. Conference papers share code significantly more often than other PubMed papers. (b) Rate of code sharing by topic: Biomedicine papers share code slightly more often. (c) Rate of code sharing by affiliation: Industry shares slightly less than others. (d) Distribution of citation counts for papers with and without code over time: Papers that share code generally have at least the same number of citations or more each year. Citation counts are limited to 50 or below to better visualize typical citation numbers beyond outlier papers. (e) Cumulative distribution of code sharing vs. no code sharing: Papers that share code overall have more citations, regardless of outlier status.}
    \label{fig:code_trends}
\end{figure}
Proprietary code in AI4H significantly hinders reproducibility. Many models use undisclosed proprietary algorithms, compromising transparency and patient safety in healthcare. A critical review \cite{Haibe-Kains2020_rebuttal_to_Google_nature_paper_with_no_code_no_data} of Google's breast cancer detection paper \cite{McKinney2020_google_breast_cancer_detection} highlights this issue, noting that insufficient details and lack of code make the high-impact work irreproducible.

To assess code-sharing trends, we tracked the presence of GitHub, Zenodo, GitLab, or Colab links within the main text as an upper-bound proxy of technical reproducibility, acknowledging that linked code may not always be complete, runnable, correctly implemented, or related to the described method. 

Figure \ref{fig:code_trends}(a) reveals increased code-sharing over time for conference venues but little change for PubMed articles (less than 20\% with code repository in their papers). CHIL and ML4H demonstrate higher rates of code mentions than MLHC, possibly due to stricter sharing requirements. However, at least 58\% of recent MLHC papers still do not share code within their main text, while PubMed's substantially lower code-sharing rate indicates that technical reproducibility remains a significant concern in AI4H. 

Figure \ref{fig:code_trends}(b-c) shows that most papers, regardless of topic or affiliation, do not share their code. Notably, industry affiliations share code slightly less often than mixed or academic affiliations. Interestingly, AI papers related to biomedicine (such as those focused on genomics) tend to share code more, while electronic health record-related (EHR) papers share code the least. 

Figure \ref{fig:code_trends}(d-e) confirmed that sharing code, similar to mentioning datasets, correlates with higher citation counts in the future. However, the significant overlap in these distributions indicates that individual academics may not notice a meaningful difference, even though a statistically significant difference exists (t-test $p < 0.05$) in Figure \ref{fig:code_trends}(e).

\subsection{Lack of Standardization in Data Processing Pipelines}
The limited standardized data processing practices in AI4H significantly impede reproducibility. Healthcare data, especially from EHRs, demands extensive preprocessing, including cohort selection, missing data management, feature normalization, and categorical variable encoding. Without a standard framework, research teams employ diverse methods, leading to substantial variability in data processing \cite{wornow2023shaky_foundations}. This variability affects model performance, complicating cross-study comparisons and result replication. 

Furthermore, improper data handling can introduce leakage, resulting in overly optimistic model performance. 
For example, studies by Kapoor et al. \cite{kapoor2022leakagereproducibilitycrisismlbased}, Vandewiele et al. \cite{Vandewiele_2021_preterm}, and Roberts et al. \cite{Roberts2021_CovidPitfalls} reveal data leakage issues in pre-term birth prediction and COVID-19 detection models, leading to inflated performance metrics. Standardizing data processing practices is, therefore, crucial for enhancing reproducibility and ensuring the reliability of AI4H models.

To promote standardization in data processing, the OHDSI community has adopted a common data model known as OMOP-CDM~\cite{noauthor_undated-mg}. This model serves as a foundation for analytic pipelines and has gained significant traction in health informatics communities. Similarly, the Medical Event Data Standard (MEDS) is another initiative aimed at standardizing data for applications that are compatible with machine learning~\cite{arnrich2024medical}. However, we note that the adoption of these standards is still incomplete.

\subsection{Error Rates of Noisy Analysis}
To validate our automated keyword search analysis, we manually examined 30 papers. Our code sharing detection achieved 87\% accuracy, while public dataset detection showed 77\% accuracy with 90\% precision (Table \ref{tab:manual_validation}). While our code sharing estimates align with manual validation, we likely underestimate public dataset usage due to the increase in availability of public health datasets beyond well-known repositories. However, given the high precision, this lower bound serves as a reliable worst-case metric. Additionally, we evaluated the medical frontier LLM's zero-shot topic classification performance. The results showed high accuracy, with discrepancies between manual and LLM classifications often representing equally valid alternative categorizations.

\begin{table}[h!]
\centering
\resizebox{\linewidth}{!}{%
\begin{tabular}{l*{6}{c}}
\toprule
\textbf{Category} & \textbf{Accuracy} & \textbf{Precision} & \textbf{Recall} & \textbf{F1 Score} & \textbf{Estimated Rate} & \textbf{True Rate} \\
\midrule
Public Dataset Usage & 0.77 & 0.90 & 0.60 & 0.72 & 0.33 & 0.50 \\
Code Sharing & 0.87 & 0.71 & 0.71 & 0.71 & 0.23 & 0.23 \\
Topic Classification & 0.93 & 0.94 & 0.93 & 0.93 & - & - \\
\bottomrule
\end{tabular}%
}
\caption{Manual validation of our automated approach on 30 randomly sampled papers, examining code sharing, public dataset usage, and topic classification. Comparing true rates from manual review against estimated rates from keyword searches, we found exact agreement between methods for code sharing (23\% both estimated and true), while public dataset usage was significantly underestimated (33\% estimated vs 50\% true). Topic classification performed exceptionally well, while code sharing detection was reliable but conservative, and public dataset detection showed high precision but missed some instances due to varied datasets not covered in the PapersWithCode API or our initial pool of popular datasets.}
\label{tab:manual_validation}
\end{table}

\section{Existing Solutions Towards Improving Reproducibility} \label{apd: existing solutions explored}
To better contextualize the role of open source software and benchmarks, we briefly explore various existing reproducibility solutions and their respective trade-offs below. While each have their own merits, they all have drawbacks that prevent widespread adoption from the AI4H research community.

\subsection{Guidelines and Standards}
Numerous reproducibility guidelines and standards have been proposed to improve research reproducibility in healthcare \cite{kapoor2022leakagereproducibilitycrisismlbased, CLAIM_Checklist_doi:10.1148/ryai.2020200029, Collinse078378_tripod, 10.1093/jamia/ocaa088_minimar, RAHROOH2024104551_automate_meta_data_checklist_reproducibility}. These checklists serve multiple purposes, including guiding authors in sharing essential information, preventing data contamination, improving the reproducibility of shared work, and providing scoring and ranking metrics to measure a paper's reproducibility \cite{pineau2020improvingreproducibilitymachinelearning, heil2021reproducibility_gold_bronze_silver}.

While useful as preventative measures and guidelines, checklists have notable limitations. They do not directly aid in implementing or validating methods, and users may input faulty information. Furthermore, studies by \cite{kapoor2022leakagereproducibilitycrisismlbased} and \cite{Edin_2023_med_coding_repl} show that many subsequent works reuse dataset preprocessing code from earlier studies that may not have undergone rigorous reproducibility audits.

Consequently, a significant body of work still requires manual auditing and validation. This task is becoming increasingly challenging as AI-related health publications have more than doubled between 2018 and 2021 \cite{JIMMA2023100041Bibliometric}. Our survey of released PubMed papers in Figure \ref{fig:scraping} supports this finding. As such, the next crucial step is to make it easier for researchers to adhere to reproducibility standards while ensuring thorough validation of existing work.

\subsection{Automation and Tooling} A range of technical tools and platforms has been developed to support reproducibility efforts, aiming to streamline workflows and enhance transparency. These tools address various aspects of the research process, from experiment tracking to model releasing. Tools like Hydra and OmegaConf enable consistent and scalable management of experimental configurations. Platforms such as Hugging Face allows efficient sharing of model checkpoints. Platforms like TensorBoard, Weights \& Biases, and Neptune provide easier management of experiments.

While these tools significantly enhance reproducibility, they are primarily designed for general machine learning tasks and are not tailored to address the unique challenges of AI4H. They are better suited as complementary counterparts rather than direct solutions for AI4H-specific reproducibility challenges. To fully address the challenges discussed earlier, AI4H requires tools explicitly designed to cater to its unique needs.

\section{Create Open-source Software and Benchmarks for Enhancing Reproducibility in AI4H}

To address the reproducibility crisis in AI for Health (AI4H), we propose the following systematic, behavior-driven strategies. 

\subsection{Open-Source Software and Benchmarks}

The foundational step in enhancing reproducibility is making tools and standards openly accessible and visible. Promoting open science through standardized data preprocessing pipelines and diverse benchmarks highlights the importance of reproducibility. Open-source platforms enhance transparency by making code, models, and workflows publicly available, while standardized benchmarks provide consistent frameworks for evaluating model performance in clinical settings, enabling validation across studies.

For example, widely adopted libraries like PyTorch~\cite{paszke2019pytorchimperativestylehighperformance}, scikit-learn~\cite{scikit-learn}, and HuggingFace~\cite{wolf2020huggingfacestransformersstateoftheartnatural} have standardized key components of machine learning and natural language processing, thus enhancing reproducibility. 
In AI4H, open-source software like \textbf{PyHealth}~\cite{10.1145/3580305.3599178_PyHealth}, MonAI \cite{cardoso2022monaiopensourceframeworkdeep_monai}, \textbf{YAIB}~\cite{vandewater2024icubenchmarkflexiblemulticenter_YAIB},  \textbf{RENOIR}~\cite{Barberis2024_renoir}, MEDS~\cite{arnrich2024medical} and open source initiatives like OHDSI \cite{hripcsak2015observational_ohdsi} become increasingly valuable for reproducibility. OHDSI has created guidelines for standardizing the format of clinical data. Open source software surrounding the standardization of publicly formatted data should not be understated.
For instance, open datasets such as MIMIC~\cite{Johnson2023-MIMIC} are frequently utilized; over 28\% of ML4H papers in recent years have used MIMIC data. To better standardize the curation and accessibility of MIMIC data, open source software such as PyHealth \cite{10.1145/3580305.3599178_PyHealth}, MEDS~\cite{arnrich2024medical}, and YAIB \cite{vandewater2024icubenchmarkflexiblemulticenter_YAIB} aid in the data preprocessing of large public datasets for various downstream clinical tasks. Taking a step further, open source container systems like Docker \cite{Openja_2022_docker_practices, moreau2023containers} can dramatically reduce the lower-level system software compatibility barriers of running AI pipelines. Clearly, promoting the availability of these resources helps researchers align on common practices, facilitating reproducibility.

\subsection{Incentivizing Reproducibility}
To encourage the widespread adoption of reproducibility practices, we must make them appealing. Open-source tools should offer benefits that outweigh any perceived burdens. Libraries like PyTorch and HuggingFace attract users through time-saving APIs, simplifying reproducibility while providing efficiency and strong community support. Recognizing and rewarding reproducible work through special publications, awards, or formal acknowledgment can further incentivize researchers. For instance, schools such as Stanford Medicine have created award programs for those who contribute greatly towards reproducible practices to further incentivize creating accessible open source software and benchmarks \cite{stanford_sporr_rare_rewards}. Establishing systems to honor those who successfully reproduce others' work promotes a culture of verification and validation within the community.

\subsection{Lowering Barriers to Contribution}
It is crucial to simplify the adoption of reproducibility practices. Enhancing functionality of these AI4H opensource tools like PyHealth~\cite{10.1145/3580305.3599178_PyHealth}, YAIB~\cite{vandewater2024icubenchmarkflexiblemulticenter_YAIB}, and MEDS~\cite{arnrich2024medical}, improving user experience, and demonstrating their impact on research can bridge this gap.

Centralized open-source platforms and established workflows enable plug-and-play model development, allowing researchers to focus on innovative problem-solving rather than redundant preprocessing steps. Improving integration with popular programming languages like Python and expanding the range of AI4H tasks covered can make these tools more accessible to the community.

\textbf{Educational Initiatives.} Open-source software is also pivotal in teaching reproducibility in graduate-level healthcare ML courses. Institutions such as MIT \cite{mit_ocw}, Johns Hopkins \cite{jhu_ml}, Georgia Tech \cite{gatech_dl}, and University of Illinois Urbana-Champaign (UIUC) \cite{uiuc_cs598} utilize open-source tools to teach key concepts, often through project-based assignments that try to replicate published research. Incorporating reproducibility-focused projects enriches students' learning experiences and promotes reproducibility. Other non-academic educational programs exist such as the Linux Foundation's open source software development course \cite{linux_foundation_ossd} and NumFOCUS's wide array of open source educational programs \cite{numfocus}. These projects can highlight the impact of reproducibility on career development. The resulting code repositories of reproducibility case studies can showcase reproducibility for the broader community. 

\subsection{Celebrating Success in Reproducibility}
Reinforcing reproducibility practices requires making the experience rewarding. Celebrating successes—such as accurate reproductions of studies or high-quality documentation—encourages continued commitment to reproducible methods. Public recognition, badges for reproducible publications, and journal incentives can create a positive feedback loop. Organizing events like reproducibility hackathons can gamify the process, making it enjoyable and collaborative.

\textbf{Moving Cultural Norms.} Part of making reproducibility sustainable is creating a culture where reproducible practices are naturally rewarded. While the volume of AI4H publications has grown exponentially over the past decade \cite{senthil2024bibliometric_trends_AI4H}, the prevailing "publish or perish" academic environment \cite{rawat2014publish_or_perish, stupple2019reproducibility_crisis_npj_digital} often forces researchers to prioritize publication output over reproducibility efforts. This pressure frequently results in research remaining private and inaccessible. However, successful examples of fostering reproducibility exist in adjacent fields. The Natural Language Processing (NLP) community, through open source platforms like HuggingFace \cite{wolf2020huggingfacestransformersstateoftheartnatural}, has demonstrated how reducing technical barriers to sharing models and datasets can transform research culture. Despite some technical limitations, HuggingFace has successfully created an environment where the visibility and recognition gained from sharing reproducible NLP pipelines outweigh the costs of implementing reproducible practices. For large foundational models in NLP, open source chatbot arenas \cite{chiang2024chatbotarenaopenplatform} have created leaderboards where the best models are given recognition, directly incentivizing users to use and interact with these models. This success story suggests a promising path forward for AI4H: by building sufficient software infrastructure and momentum, we can lower the barriers to sharing models and benchmarking on public datasets. A more health related example is PhysioNet \cite{PhysioNet}, which has streamlined the sharing of medical datasets and become a pillar in the sharing of the dominant MIMIC \cite{Johnson2023-MIMIC} series of datasets, commonly used across the AI4H space. While this approach may not solve all reproducibility challenges, it could significantly shift the cultural attitudes toward reproducibility in the field. 

\textbf{Reproducibility Hackathons.} 
While open-source software is essential for enabling reproducibility in research, it can also serve as a powerful catalyst to promote and advocate for reproducibility by significantly reducing development time~\cite{Cokelaer2023-vz_reprohackathons, bender2024libraries_CMU_reprohackathons}. In the field of healthcare AI, enhanced tools make it easier for both students and professionals to participate in hackathons, fostering collaboration on shared datasets like MIMIC and the UK Biobank. Participants can share code and concentrate on documenting reproducible workflows, ultimately building a public repository of reproducible solutions. Offering sponsorships and prizes can enhance the experience, increase engagement, and underscore the importance of reproducibility. For example, a recent Kaggle competition on EEG classification organized by academic researchers received \$50K in sponsorship from pharmaceutical, medical device, and technology companies and attracted hundreds of submissions~\cite{UnknownUnknown-yy}.

\subsection{Enforcing Reproducible Practices}
The ultimate goal is to encourage widespread adoption of reproducible practices across the research community. As discussed in Section \ref{sec: Crisis}, while patient privacy concerns often prevent the direct replication of study results, sharing code generally poses no privacy risks. Our analysis in Figure \ref{fig:code_trends}a reveals a notable trend: conference venues that have implemented code sharing mandates have achieved substantially higher rates of code mentions in publications. In contrast, journal publications (indexed in PubMed) have not yet widely adopted such requirements, as shown in Figure \ref{fig:code_trends}a - a gap that could be addressed through policy changes. For academic researchers, who unlike their industry counterparts are not constrained by proprietary code concerns, code sharing should be a straightforward and enforceable requirement.

\subsection{\rev{Moving Targets of Reproducibility}}
\rev{As datasets and models evolve across environments and time, reproducibility remains a moving target in both healthcare systems and when adopting new models. Well-written open source software addresses this challenge by enabling researchers and practitioners to seamlessly integrate state-of-the-art models and retrain them on new data distributions—crucial for both research advancement and practical implementation.
While reported metrics may not reflect real-world performance on target patient populations, evaluating new models typically requires access to source code, especially when training details are vaguely described in the literature. Transparent public benchmarks become especially valuable as training costs grow prohibitively expensive, as with LLMs \cite{zhao2023surveylargelanguagemodels}, allowing practitioners to reference established metrics before investing in adaptation to private data.
Although sensitive patient data in healthcare settings must remain protected due to privacy regulations, publicly available code simplifies model adaptation to these privacy-sensitive target distributions. Moreover, open-source software directly addresses the challenge of shifting distributions over time. By enabling evaluation of published models on specific cohorts across time periods, it facilitates explorations in conceptual reproducibility—determining whether one model consistently outperforms another across changing domains. Ultimately, methods lacking publicly available, well-documented code are frequently overlooked or avoided by healthcare practitioners, regardless of their potential benefits, as implementation risks outweigh theoretical advantages.}

\subsection{The Future of Reproducibility with Open Source Software and AI Agents}
While Large Language Models (LLMs) alone may not be reliable judges of reproducibility \cite{10.1145/3641525.3663629_LLM_reproduc}, they demonstrate remarkable capabilities when integrated with tools and programmatic APIs as agents \cite{li2024reviewprominentparadigmsllmbased_tool_agents, yang2024llmwizardcodewand}. These AI agents have shown proficiency in diverse healthcare tasks, from reasoning over electronic health records \cite{shi2024ehragentcodeempowerslarge_EHR_agent} to conducting data analysis \cite{wuehrflow} using existing open source software tools. The development of more robust open source software tools for healthcare model deployment and public dataset access could significantly enhance these LLM agents' capabilities. Such tools would enable AI agents to more effectively reproduce and prototype AI pipelines for healthcare data. This technological advancement could provide a scalable solution to the growing challenge of reproducing results from the exponentially increasing number of AI4H publications.

\section{Conclusion}
The reproducibility crisis in AI for Healthcare urgently demands attention due to its direct impact on human lives. Our analysis reveals significant challenges: 74\% of AI4H papers analyzed in this paper rely on private datasets or do not share code, and data processing lacks standardization. Although rare, papers that use public datasets and share code tend to receive more citations, indicating that embracing reproducible practices correlates with greater research impact.

The general machine learning community has embraced open-source practices for over a decade, and it is crucial for AI in health (AI4H) to do the same due to the high stakes involved. By prioritizing open-source development, we can enhance reproducibility, accelerate progress, and enable researchers to concentrate on innovation instead of repeatedly developing data processing steps. This shift will foster trust in AI healthcare solutions by promoting transparency, which is essential for safe and effective deployment in clinical settings. The future of AI in healthcare relies not only on breakthroughs but also on our ability to validate, reproduce, and implement innovations safely in real-world practice.

\bibliography{Bibliography} 
\pagebreak
\appendix
\section{Appendix}
\subsection{Paper Pipeline and Analysis Details}\label{apd: Methodology Details}
We outline the many steps we take to aggregate and process the data used in the visualizations above in Figure \ref{fig:scraping}. We also provide an error analysis of our keyword-based approach towards gauging the state of reproducibility. \rev{Our existing data and implementation is located at \url{https://github.com/sunlabuiuc/reproai4h}
.}

\subsubsection{Paper Collection Details} \label{apd:collection}
To analyze recent trends in AI4H research (2018-2024), we examined papers from three prominent healthcare-focused AI conferences: Machine Learning for Health (ML4H), Conference on Health, Inference, and Learning (CHIL), and Machine Learning for Healthcare (MLHC). This timeframe continues from \citep{mcdermott2019reproducibilitymachinelearninghealthcrisis}'s reproducibility study. We selected these conferences due to their accessible, web-scrapable proceedings, unlike other venues such as AMIA. From an initial collection of 635 papers, we developed an automated pipeline to extract metadata. Since PDF text lacks explicit metadata labels, we extracted titles by identifying all text before the first person string using NER language models, and captured author details from text between the title and abstract. We then extracted abstracts by locating text preceding the introduction section header and employed a quantized 70B Llama 3.1 model to clean titles with the prompt shown in Figure \ref{fig:title-cleaning-prompt} and extract email addresses from author details with the prompt shown in Figure \ref{fig:email-extraction-prompt}. We then used various API services to gather citation data for each paper in Section \ref{apd:citation_count}. After filtering out papers with incomplete metadata, missing citation data, and non-research content (such as proceedings summaries, abstracts, and demos), our final dataset comprised 528 papers.

\begin{figure}[h!]
    \begin{tcolorbox}[
        colback=white,
        colframe=black,
        arc=0mm,
        boxrule=0.5pt
    ]
    \small
    You are an assistant specialized in cleaning and standardizing academic paper titles. Your task is to take a given title and improve its formatting, spacing, and consistency. Follow these rules:
    
    \textbf{1. Correct spacing:}
    \begin{itemize}
        \item Ensure single spaces between words.
        \item Remove extra spaces before or after hyphens.
        \item Add spaces after colons and semicolons.
    \end{itemize}
    
    \textbf{2. Hyphenation:}
    \begin{itemize}
        \item Use hyphens consistently in compound terms (e.g., ``Multi-Scale'' not ``Multi Scale'' or ``MultiScale'').
        \item Correct common hyphenation errors in technical terms (e.g., ``Pre-processing'' not ``Preprocessing'').
    \end{itemize}
    
    \textbf{3. Capitalization:}
    \begin{itemize}
        \item Use title case: Capitalize the first letter of each major word.
        \item Do not capitalize articles (a, an, the), coordinating conjunctions (and, but, for, or, nor), or prepositions unless they start the title.
        \item Always capitalize the first and last words of the title and subtitle.
    \end{itemize}
    
    \textbf{4. Acronyms and initialisms:}
    \begin{itemize}
        \item Remove spaces between letters in acronyms (e.g., ``CNN'' not ``C N N'').
        \item Ensure correct formatting of technical acronyms (e.g., ``U-Net'' not ``UNet'' or ``U Net'').
    \end{itemize}
    
    \textbf{5. Special characters:}
    \begin{itemize}
        \item Correct the use of special characters like hyphens (-), en dashes (--), and em dashes (---).
        \item Ensure proper use of quotation marks and apostrophes.
    \end{itemize}
    
    \textbf{6. Consistency:}
    \begin{itemize}
        \item Maintain consistent formatting throughout the title.
        \item Ensure that similar terms or concepts are formatted the same way.
    \end{itemize}
    
    \textbf{7. Grammar and spelling:}
    \begin{itemize}
        \item Correct any obvious spelling errors.
        \item Ensure proper grammatical structure.
    \end{itemize}
    
    \textbf{8. No Authors:} If the title contains any author names, emails, or affiliations, remove them.
    
    When given a title, apply these rules to clean and standardize it. Provide the corrected title without additional commentary unless there are ambiguities or decisions that require explanation.
    
    \texttt{Title to clean: \{title\}}\\
    \texttt{Cleaned title:}
    \end{tcolorbox}
    \caption{Prompt for standardizing noisily extracted conference paper titles.}
    \label{fig:title-cleaning-prompt}
\end{figure}

\begin{figure}[h!]
    \centering
    \begin{tcolorbox}
Extract all email addresses from the following text.
Clean the extracted email addresses by removing any unnecessary characters 
or formatting issues.
Output only the cleaned email addresses, one per line.
Text: \{text\}
Cleaned and extracted email addresses:
    \end{tcolorbox}
    \caption{Prompt template for extracting and cleaning email addresses from author details using a large language model. The \{text\} is the extracted author detail text from the noisy pdf files.}
    \label{fig:email-extraction-prompt}
\end{figure}

We used Python's Entrez API to query PubMed with a search string generated through PubMed's advanced search interface (https://pubmed.ncbi.nlm.nih.gov/), collecting a list of PMIDs identifying AI4H papers. Our query, referenced in Figure \ref{fig:pubmed_query}, targeted papers containing AI-related terms (e.g., deep learning, machine learning, artificial intelligence) and clinical terms (e.g., electronic health record, electronic medical record, healthcare) while excluding non-research articles such as surveys and peer reviews. Using the resulting PMIDs, we extracted paper content as XML files through the BioC API \cite{gusenbauer2020academic_pmid_cite}. From the initial 2,404 PMIDs identified, we retrieved 2,082 XML papers, with the difference due to limited open access availability. The BioC-formatted XML files included well-structured paper text and metadata, eliminating the need for additional cleaning.

\definecolor{codegreen}{rgb}{0,0.6,0}
\definecolor{codegray}{rgb}{0.5,0.5,0.5}
\definecolor{codepurple}{rgb}{0.58,0,0.82}
\definecolor{backcolour}{rgb}{0.95,0.95,0.92}

\fvset{
  commandchars=\\\{\},
  numbers=left,
  numbersep=5pt,
  frame=single,
  framesep=3mm,
  rulecolor=\color{codegray},
  fontsize=\footnotesize,
  baselinestretch=1.2,
  breaklines=true,
  breakanywhere=true,
  tabsize=2
}

\begin{figure}[h!]
\centering
\begin{Verbatim}[breaklines=true, breakanywhere=true, commandchars=\\\{\}]
query = '((("machine learning"[MeSH Terms] OR ("machine"[All Fields] AND \
"learning"[All Fields]) OR "machine learning"[All Fields] OR ("deep learning"\
[MeSH Terms] OR ("deep"[All Fields] AND "learning"[All Fields]) OR "deep \
learning"[All Fields]) OR ("artificial intelligence"[MeSH Terms] OR \
("artificial"[All Fields] AND "intelligence"[All Fields]) OR "artificial \
intelligence"[All Fields])) AND ("electronic health records"[MeSH Terms] OR \
("electronic"[All Fields] AND "health"[All Fields] AND "records"[All Fields]) \
OR "electronic health records"[All Fields] OR ("electronic"[All Fields] AND \
"health"[All Fields] AND "record"[All Fields]) OR "electronic health record"\
[All Fields] OR ("ethics hum res"[Journal] OR "environ hist rev"[Journal] OR \
"ehr"[All Fields]) OR ("empir musicol rev"[Journal] OR "emr"[All Fields]) OR \
("electronic health records"[MeSH Terms] OR ("electronic"[All Fields] AND \
"health"[All Fields] AND "records"[All Fields]) OR "electronic health records"\
[All Fields] OR ("electronic"[All Fields] AND "medical"[All Fields] AND \
"record"[All Fields]) OR "electronic medical record"[All Fields]) OR \
("delivery of health care"[MeSH Terms] OR ("delivery"[All Fields] AND "health"\
[All Fields] AND "care"[All Fields]) OR "delivery of health care"[All Fields] \
OR "healthcare"[All Fields] OR "healthcare s"[All Fields] OR "healthcares"\
[All Fields]))) NOT ("survey s"[All Fields] OR "surveyed"[All Fields] OR \
"surveying"[All Fields] OR "surveys and questionnaires"[MeSH Terms] OR \
("surveys"[All Fields] AND "questionnaires"[All Fields]) OR "surveys and \
questionnaires"[All Fields] OR "survey"[All Fields] OR "surveys"[All Fields] \
OR ("review"[Publication Type] OR "review literature as topic"[MeSH Terms] OR \
"review"[All Fields]) OR ("perspective"[All Fields] OR "perspective s"\
[All Fields] OR "perspectives"[All Fields]) OR ("comment"[Publication Type] \
OR "commentary"[All Fields]) OR ("comment"[Publication Type] OR "viewpoint"\
[All Fields]) OR ("editorial"[Publication Type] OR "editorial"[All Fields]) \
OR (("letter"[Publication Type] OR "correspondence as topic"[MeSH Terms] OR \
"letter"[All Fields]) AND to the[Author] AND ("editor"[All Fields] OR \
"editor s"[All Fields] OR "editors"[All Fields])) OR ("letter"\
[Publication Type] OR "correspondence as topic"[MeSH Terms] OR "correspondence"\
[All Fields]) OR ("guideline"[Publication Type] OR "guidelines as topic"\
[MeSH Terms] OR "guideline"[All Fields]) OR (("patient positioning"\
[MeSH Terms] OR ("patient"[All Fields] AND "positioning"[All Fields]) OR \
"patient positioning"[All Fields] OR "positioning"[All Fields] OR "position"\
[All Fields] OR "position s"[All Fields] OR "positional"[All Fields] OR \
"positioned"[All Fields] OR "positionings"[All Fields] OR "positions"\
[All Fields]) AND ("paper"[MeSH Terms] OR "paper"[All Fields] OR "papers"\
[All Fields] OR "paper s"[All Fields])) OR ("consens statement"[Journal] OR \
("consensus"[All Fields] AND "statement"[All Fields]) OR "consensus statement"\
[All Fields]) OR ("systematic review"[Publication Type] OR "systematic reviews \
as topic"[MeSH Terms] OR "systematic review"[All Fields]) OR ("meta analysis"\
[Publication Type] OR "meta analysis as topic"[MeSH Terms] OR "meta analysis"\
[All Fields]) OR ("case reports"[Publication Type] OR "case report"\
[All Fields]) OR ("review"[Publication Type] OR "review literature as topic"\
[MeSH Terms] OR "literature review"[All Fields]) OR ("educability"[All Fields] \
OR "educable"[All Fields] OR "educates"[All Fields] OR "education"\
[MeSH Subheading] OR "education"[All Fields] OR "educational status"\
[MeSH Terms] OR ("educational"[All Fields] AND "status"[All Fields]) OR \
"educational status"[All Fields] OR "education"[MeSH Terms] OR "education s"\
[All Fields] OR "educational"[All Fields] OR "educative"[All Fields] OR \
"educator"[All Fields] OR "educator s"[All Fields] OR "educators"[All Fields] \
OR "teaching"[MeSH Terms] OR "teaching"[All Fields] OR "educate"[All Fields] \
OR "educated"[All Fields] OR "educating"[All Fields] OR "educations"\
[All Fields]))) AND ((ffrft[Filter]) AND (classicalarticle[Filter] OR \
clinicalstudy[Filter] OR clinicaltrial[Filter] OR clinicaltrialphasei[Filter] \
OR clinicaltrialphaseii[Filter] OR clinicaltrialphaseiii[Filter] OR \
clinicaltrialphaseiv[Filter] OR dataset[Filter] OR governmentpublication\
[Filter] OR preprint[Filter] OR randomizedcontrolledtrial[Filter] OR \
researchsupportamericanrecoveryandreinvestmentact[Filter] OR \
researchsupportnihextramural[Filter] OR researchsupportnihintramural[Filter] \
OR researchsupportnonusgovt[Filter] OR researchsupportusgovtnonphs[Filter] OR \
researchsupportusgovtphs[Filter] OR researchsupportusgovernment[Filter] OR \
technicalreport[Filter] OR validationstudy[Filter])) AND pubmed pmc open \
access[filter]'
\end{Verbatim}
\caption{PubMed Query for Entrez API to retrieve papers containing AI-related terms (e.g., deep learning, machine learning, artificial intelligence) and clinical terms (e.g., electronic health record, electronic medical record, healthcare) while excluding non-research articles such as surveys and peer reviews. }
\label{fig:pubmed_query}
\end{figure}

Overall, we start with an initial total number of 3,039 papers and filter down to 2,610 papers as originally shown in Figure \ref{fig:scraping}. 

\subsubsection{Citation Count Extraction} \label{apd:citation_count}
For citation analysis, we employed multiple services to ensure comprehensive coverage. For PubMed papers, we used PMIDcite \cite{gusenbauer2020academic_pmid_cite} to extract citation counts directly. For conference papers, which are not indexed by PMIDcite, we first queried Semantic Scholar \cite{Kinney2023TheSS_Semantic_scholar}, then used SerpAPI's \cite{serpapi2024} Google Scholar service for papers not found on Semantic Scholar. Papers without citation data from any of these services were excluded from our analysis. We note that while PMIDcite is freely accessible, SerpAPI has a limit of 100 free API calls, and Semantic Scholar requires an access request.

\subsubsection{Affiliation Extraction}
For affiliation analysis, conference papers typically include both emails and author affiliations, while PubMed data requires additional processing. We used PubMed's MEDLINE API \citep{gusenbauer2020academic_pmid_cite} to retrieve author affiliations for each PMID. Due to data availability limitations, MEDLINE entries lack email addresses, and conference PDF extractions may have incomplete email data. We developed a two-pronged approach. When available, we used email suffixes to classify affiliations as industry, academic, or mixed. Otherwise, we relied on affiliation keywords as detailed in Table \ref{tab:aff_classification_criteria}. For papers with both email and affiliation data, we used the resolution algorithm in Figure \ref{fig:affiliation-resolution} to reconcile any discrepancies and maximize valid affiliation coverage with our keyword approach.

\begin{figure}[h!]
\begin{verbatim}
if email_category == author_category:
    return email_category
elif 'mixed' in [email_category, author_category]:
    return 'mixed'
elif 'academic' in [email_category, author_category] and \
     'industry' in [email_category, author_category]:
    return 'mixed'
elif email_category == 'other':
    return author_category
elif author_category == 'other':
    return email_category
else:
    return 'mixed'
\end{verbatim}
\caption{Algorithm for resolving author affiliation categories based on email and author information.}
\label{fig:affiliation-resolution}
\end{figure}

\begin{table}
\centering
\begin{tabular}{|l|p{0.25\textwidth}|p{0.5\textwidth}|}
\hline
\textbf{Category} & \textbf{Email Domain Suffixes} & \textbf{Affiliation Keywords} \\
\hline
Industry & \begin{itemize}
\item .com
\item .co
\item .org
\item .io
\end{itemize} & 
\begin{itemize}
\item google, ibm, iqvia, microsoft, apple, amazon
\item facebook, intel, nvidia, deargen, riken
\item corporation, inc, llc, ltd, gmbh, company
\item biotech, analytics, corp, tno
\item leo innovation lab, widex, axispoint
\item optum, tencent, analytix
\end{itemize} \\
\hline
Academic & \begin{itemize}
\item .edu
\item .ac
\item .eth.ch
\item .cispa.de
\item dkfz-heidelberg.de
\end{itemize} & 
\begin{itemize}
\item university, school, college, institute of technology
\item polytechnic, instituto superior tecnico
\item universit, hospital, medical center
\item health science, academia, faculty
\item department of, school of medicine
\item research institute, national institute
\item cancer center, clinic
\item basque center for applied mathematics
\item association, fau erlangen-nurnberg
\item national, isi foundation, institute for health
\item amsterdam umc, iust, mayo clinic, NIH
\item .ac.uk, universidad
\end{itemize} \\
\hline
\end{tabular}
\caption{Keyword Classification Criteria for Industry and Academic Affiliations. Papers were classified as academic, industry, or mixed based on author affiliations. Paper author details with only academic keywords were labeled academic, only industry keywords as industry, and those containing both as mixed affiliations. }
\label{tab:aff_classification_criteria}
\end{table}

\subsubsection{Topic Classification}
We performed topic classification using the medically fine-tuned, quantized Llama 3 OpenBioLLM-70B model \citep{jain2024llama3openbio}, leveraging its healthcare domain knowledge. Using the prompt shown in Figure \ref{fig:classification-prompt}, we categorized each paper's title and abstract into one of four categories: Electronic Health Records (EHR), Clinical Images, Biomedicine, and Biosignals.

\begin{figure}[h!]
    \begin{center}
    \fbox{\parbox{0.95\textwidth}{
    \small
    Given the following title and abstract, classify the text into one of the main categories based on the provided guidelines. If multiple categories seem applicable, choose the most relevant one.
    
    \textbf{Title:} Predictive Modeling of Patient Outcomes Using EHR Data
    
    \textbf{Abstract:} This study leverages machine learning techniques to analyze electronic health records (EHR) for predicting patient outcomes in a large hospital network. We explore various data mining approaches to extract meaningful patterns from diverse clinical data, including diagnoses, treatments, and administrative information, to develop a robust predictive model for patient prognosis and resource allocation.
    
    \textbf{Main Categories:}
    \begin{enumerate}
        \item \textbf{Clinical Images:} Visual medical data such as X-rays, MRIs, CT scans, or other imaging techniques used for diagnosis or monitoring.
        \item \textbf{Biosignals:} Electrical or chemical signals from the body, such as heart rate, brain activity, or muscle contractions.
        \item \textbf{Biomedicine:} Molecular and cellular level studies, including genetics, protein analysis, or metabolic processes.
        \item \textbf{E.H.R (Electronic Health Records):} Digital versions of patients' medical history, including diagnoses, treatments, and administrative data.
    \end{enumerate}
    
    \textbf{Key Guidelines for Classification:}
    \begin{enumerate}
        \item \textbf{Data Type:} Consider the primary type of data being analyzed or discussed (e.g., images, electrical signals, molecular data, or patient records).
        \item \textbf{Research Focus:} Identify the main area of study (e.g., diagnostic imaging, physiological monitoring, molecular biology, or healthcare management).
        \item \textbf{Methodology:} Look for specific techniques or tools mentioned (e.g., image processing, signal analysis, sequencing, or data mining).
        \item \textbf{Application:} Consider the intended use of the research (e.g., diagnosis, treatment planning, drug discovery, or health system optimization).
        \item \textbf{Scale:} Note the scale of the study (e.g., individual organs, whole-body systems, molecular level, or population-level data).
    \end{enumerate}
    
    Please provide your classification in the following format:\\
    \texttt{Category: [Main Category]}
    
    \texttt{Classification:}
    }}
    \caption{Prompt for classifying medical research papers into 4 topics: Electronic Health Records (EHR), Clinical Images, Biomedicine, and Biosignals. }
    \label{fig:classification-prompt}
    \end{center}
\end{figure}

\subsubsection{Code Sharing Measurement Details} \label{apd:code_sharing}
While measuring code sharing has inherent limitations—shared repositories may not directly relate to the presented methods or may contain non-functional code—we developed an automated approach to gauge code availability. We identified papers with shared code by detecting mentions of common code-sharing platforms, specifically GitHub, Zenodo, Colab, and GitLab. Any paper containing at least one reference to these platforms was classified as having "shared code." We ignore any text from the references. 

\subsubsection{Public Dataset Measurement Details}\label{apd:public_dataset}
To assess public dataset usage, we developed a proxy measure focusing on widely-cited clinical and biomedical datasets. Our analysis searched for mentions of: MIMIC \citep{Johnson2023-MIMIC}, eICU \citep{eICU}, UK BioBank \citep{Bycroft2018UKBioBank}, Chest X-Ray14 \citep{ge2018chestxraysclassificationmultilabel}, The Cancer Genome Atlas Program (TCGA) \citep{Cancer_Genome_Atlas_Research_Network2013-zd}, Genomic Data Commons (GDC) \citep{Grossman2016_GDC}, SEER \citep{SEER2024}, OASIS \citep{OASIS1}, PhysioNet \citep{PhysioNet}, ADNI \citep{Shi2021-ay_ADNI}, and Temple University Hospital EEG Corpus \citep{10.3389/fnins.2016.00196_TUH_DATA}. To ensure comprehensive coverage, we supplemented our search with the Papers With Code API. Papers mentioning any of these datasets or their abbreviations were classified as using public datasets. We ignore any text from the references. We showcase the keywords that we used to manually search for dataset mentions in Table \ref{tab:dataset_terms}. 

\begin{table}[htbp]
    \centering
    \begin{tabular}{p{0.3\textwidth}p{0.6\textwidth}}
    \toprule
    \textbf{Dataset} & \textbf{Search Terms} \\
    \midrule
    MIMIC & Medical Information Mart for Intensive Care \\
    eICU & eICU Collaborative Research Database \\
    UK Biobank & UK Biobank \\
    Chest X-Ray14 & Chest X-Ray14, NIH Chest X-ray \\
    ADNI & ADNI, Alzheimer's Disease Neuroimaging Initiative \\
    PhysioNet & PhysioNet \\
    OASIS & OASIS, Open Access Series of Imaging Studies \\
    TCGA & TCGA, The Cancer Genome Atlas Program \\
    GDC & GDC, Genomic Data Commons \\
    SEER & SEER, Surveilance Epidemiology and End Results \\
    TUH EEG Corpus & TUH EEG Corpus, TUEG \\
    TUH Abnormal EEG Corpus & TUH Abnormal EEG Corpus, TUAB \\
    TUH EEG Artifact Corpus & TUH EEG Artifact Corpus, TUAR \\
    TUH EEG Epilepsy Corpus & TUH EEG Epilepsy Corpus, TUEP \\
    TUH EEG Events Corpus & TUH EEG Events Corpus, TUEV \\
    TUH EEG Seizure Corpus & TUH EEG Seizure Corpus, TUSV \\
    TUH EEG Slowing Corpus & TUH EEG Slowing Corpus, TUSL \\
    \bottomrule
    \end{tabular}
    \caption{Dataset names and their corresponding search terms used in the analysis.}
    \label{tab:dataset_terms}
\end{table}

\subsubsection{Additional Analysis on The Relationship Between Reproducibility and Citations}
\begin{figure}
    \centering
    \includegraphics[width=1.0\linewidth]{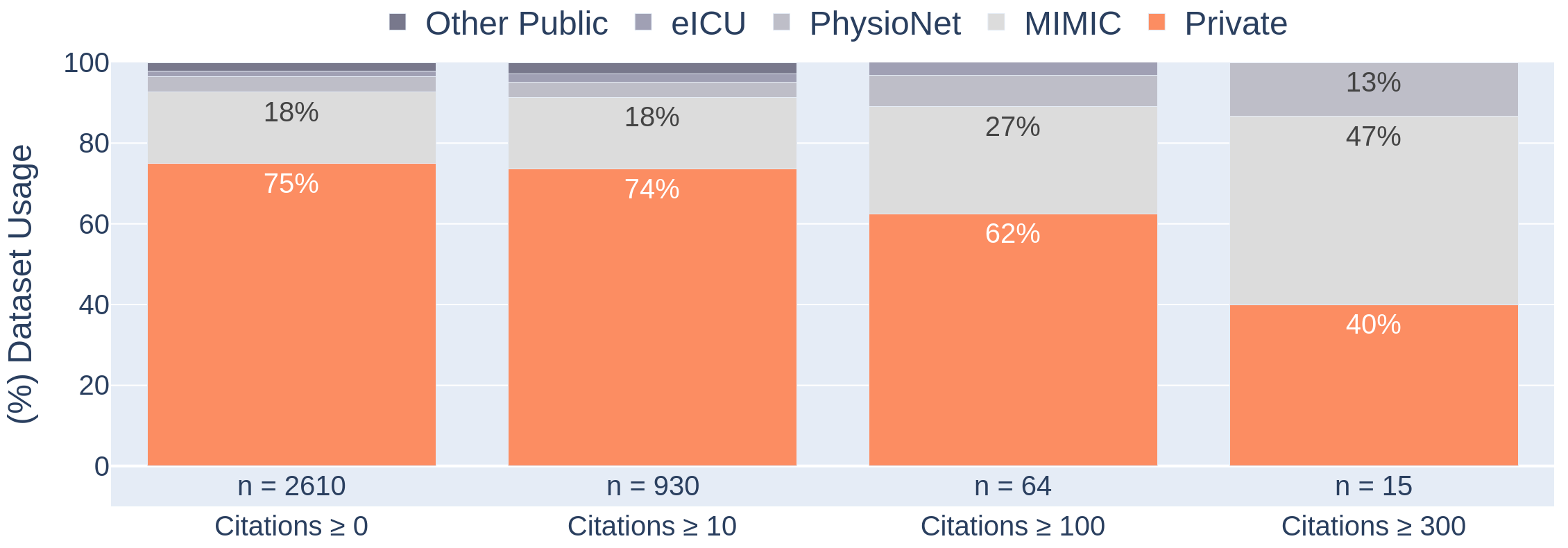}
    \caption{Papers with higher citation counts show increased usage of public datasets, as also demonstrated by the cumulative distribution plot in Figure \ref{fig:data_trends}e. This trend indicates a positive correlation between paper visibility and researchers' tendency to use public datasets. However, the majority of papers have less than 10 citations and the vast majority of such papers use private datasets, suggesting that the benefits of reproducible practices are not noticeable for the vast majority of papers.}
    \label{fig:data_citation_thresholded}
\end{figure}

\begin{figure}
    \centering
    \includegraphics[width=1.0\linewidth]{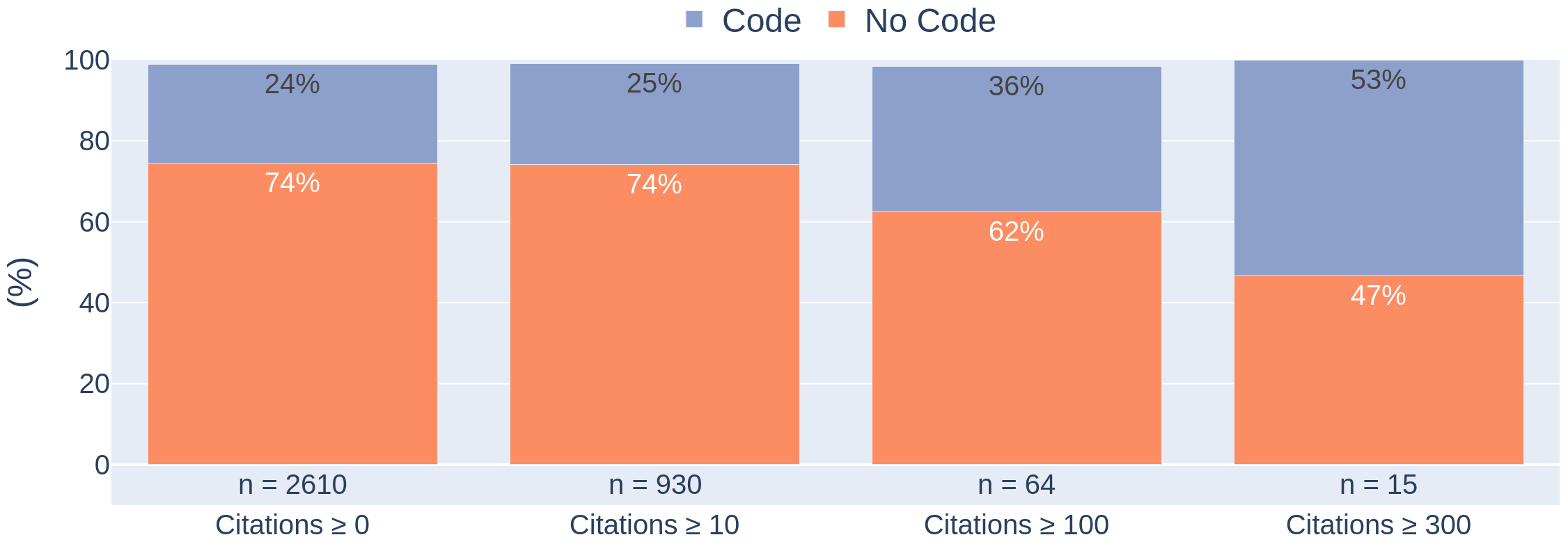}
    \caption{Figure shows the relationship between paper citations and code availability across different citation thresholds. The proportion of papers with available code increases with citation count, from 24\% for papers with any citations to 53\% for highly-cited papers ($\geq $300 citations). While most papers in the dataset (n=2,610) have relatively few citations, those with higher citation counts demonstrate improved code sharing practices. This follows the observations in the empirical distribution plot in Figure \ref{fig:code_trends}e.}
    \label{fig:code_citation_thresholded}
\end{figure}

Our analysis of the relationship between reproducibility and citation counts reveals nuanced patterns. While Figures \ref{fig:code_citation_thresholded} and \ref{fig:data_citation_thresholded} show that most papers receive relatively few citations regardless of their reproducibility practices, there is a notable trend: papers with higher citation counts are more likely to implement reproducible practices. This positive correlation between citations and reproducibility persists across different citation thresholds, which follows with the the statistical significance of the difference between reproducible and non-reproducible papers remains consistent as shown in Figures \ref{fig:code_trends} and \ref{fig:data_trends}.

\end{document}